# Unsupervised Machine Learning for Osteoporosis Diagnosis Using Singh Index Clustering on Hip Radiographs

**Vijaya Kalavakonda[1,*], Vimaladevi Madhivanan[1,*], Abhay Lal[1,*], Senthil Rithika[1], Shamala Karupusamy Subramaniam[2], Mohamed Sameer[3]**

[1]*Department of Computing Technologies, School of Computing, SRM Institute of Science and Technology, Kattankulathur, Chennai 603203, India.*

[2]*Department of Communication Technology and Network, Faculty of Computer Science and Information Technology, Universiti Putra Malaysia, Malaysia*

[3]*Department of Orthopaedics, Sri Ramachandra Medical College, Porur, Chennai 600 116. India.*

**Corresponding author Email: vijayak@srmist.edu , vimaladm@srmist.edu.in , abhaylal@icloud.com*

## Abstract

Osteoporosis, a prevalent condition among the aging population worldwide, is characterized by diminished bone mass and altered bone structure, increasing susceptibility to fractures. It poses a significant and growing global public health challenge over the next decade. Diagnosis typically involves Dual-energy X-ray absorptiometry to measure bone mineral density, yet its mass screening utility is limited. The Singh Index (SI) provides a straightforward, semi-quantitative means of osteoporosis diagnosis through plain hip radiographs, assessing trabecular patterns in the proximal femur. Although cost-effective and accessible, manual SI calculation is time-intensive and requires expertise. This study aims to automate SI identification from radiographs using machine learning algorithms. An unlabelled dataset of 838 hip X-ray images from Indian adults aged 20-70 was utilized. A custom convolutional neural network architecture was developed for feature extraction, demonstrating superior performance in cluster homogeneity and heterogeneity compared to established models. Various clustering algorithms categorized images into six SI grade clusters, with comparative analysis revealing only two clusters with high Silhouette Scores for promising classification. Further scrutiny highlighted dataset imbalance and emphasized the importance of image quality and additional clinical data availability. The study suggests augmenting X-ray images with patient clinical data and reference images, alongside image pre-processing techniques, to enhance diagnostic accuracy. Additionally, exploring semi-supervised and self-supervised learning methods may mitigate labelling challenges associated with large datasets.

***Abbreviations***

| | |
|---|---|
| ANN | Artificial neural network |
| BMD | Bone mineral density |
| CNN | Convolutional neural network |
| CT | Computed tomography |
| DL | Deep learning |
| DNN | Deep neural network |
| DXA | Dual-Energy X-ray Absorptiometry |
| KL | Kellgren-Lawrence |
| ML | Machine learning |
| SI | Singh Index |

## 1. Introduction

Osteoporosis is a metabolic condition of the musculoskeletal system caused due to low bone mass and micro-architectural deterioration, which increases bone fragility and fracture risk. This is a major health concern that increases mortality along with economic and social burden worldwide. As the global population is aging, it is expected that the osteoporosis risk increases exponentially and is in fact termed as the next pandemic. According to historical data [1], on average, 50% of women and 20% of men will be affected by osteoporotic fractures in their lifetime. Worldwide statistics show that over the age of 50, 6.3% of men are affected by the disease and in women it is about 21.2% [2]. Shatrugna et al. [3] investigated the prevalence of osteopenia and osteoporosis in women aged 30-60 years from low-income. Their study indicated that the nutritional status of women significantly influences bone parameters. Osteoporosis is asymptomatic and will not be identified until the patient sustains a fracture, after which the diagnosis is made by measuring bone mineral density using dual-energy x-ray absorptiometry (DXA). This machine measures the number of X-rays absorbed by tissues and converts the bone density information into a T-score and Z score based on an arithmetical algorithm comparing similarly aged Caucasian patient dataset. The World Health Organization has defined the usage of T-score to classify and define bone mineral density (BMD) measurements [4]. Depending on the T-score, a person is:

- Within the normal range: A T-score within 1 standard deviation above or below the young adult mean (±1).

- Osteopenia: A T-score falling between 1 and 2.5 below the young adult mean (-1 to -2.5).
- Osteoporosis: A T-score exceeding 2.5 below the young adult mean (-2.5 to -4.0).

Though it is a quick, painless, and non- invasive test, it is costly and not readily available everywhere. Another major disadvantage of using the DXA scan is discrepancies in the measurements. The results may vary for different manufacturers, models, modes of operation, software versions, operator errors, and so on. The cost of the instrument is very high, which makes its availability difficult in rural areas, especially in developing and under-developed countries [5]. Hence, there is a need to find a more affordable and less error-prone method for this osteoporosis prediction problem.

The Singh Index (SI) is a simple, semi-quantitative evaluation tool used in the diagnosis of osteoporosis with plain radiographs [6]. It is based on the trabecular pattern of the proximal femur. Principal compression, secondary compression, primary tensile, secondary tensile, and intertrochanteric are trabecular types in the proximal part of the femur. Hip fracture risk is frequently stated as a relative risk. Kanis et al. [7] assessed the usefulness of the World Health Organization diagnostic criteria in determining the relative risks of hip fracture in men and women. As osteoporosis progresses in a patient, these trabeculae get thinner and eventually disappear. Hence, based on the visibility of these trabecular types, the SI index classifies the X-ray image of the femur bone into six grades of the SI index. The image will be graded as:

- Grade 1: Indicates visibility of only a thin principal compression trabecula.
- Grade 2: Implies visibility of principal compression trabeculae alongside resorbed other trabeculae.
- Grade 3: Represents thinned principal tensile trabeculae with a breakage in continuity.
- Grade 4: Denotes thinned principal tensile trabeculae without loss of continuity.
- Grade 5: Signifies visibility of principal tensile and compression trabeculae, along with the presence of a Ward triangle.
- Grade 6: Indicates visibility of all trabeculae, each maintaining normal thickness.

If the calculated grade is below 4, it confirms that the person is having osteoporosis and the same is indicated in Figure 1 [8]. This method of using SI is available for use in mass screening due to the availability of plain films that can be obtained at most outpatient clinics. The effectiveness of SI in assessing bone strength and structural integrity is proven in many studies and can be used as an effective screening tool that can help patients at high-risk.

The classification of SI is done manually using X-ray images. This work aims to develop a method that could be used to automate the classification process by using deep learning techniques. It intends to reduce the inaccuracies that may arise in the manual process and can prove to be a much faster and cost-effective approach. The recent advancements in artificial intelligence and the application of machine learning algorithms in medicine and biology have proven to provide satisfactory results in such prediction systems. Despite the increased interest in applying deep learning models in medical imaging, many challenges need to be overcome due to the unavailability of labelled data. In this research work, the applicability and the effectiveness of clustering algorithms, namely K-Means, are studied.

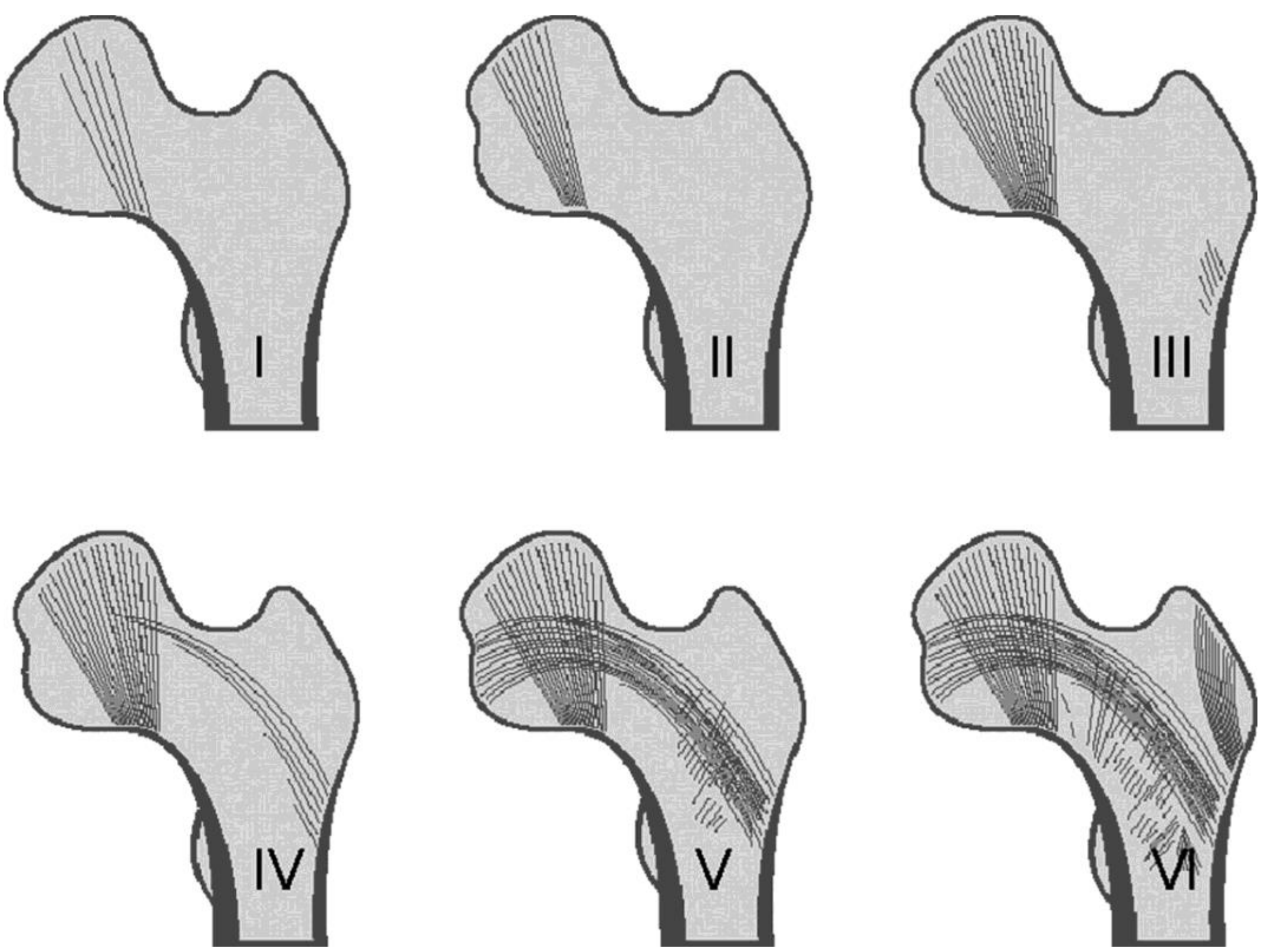

**Figure 1.** SI Grades for Osteoporosis (Reuse permission was obtained from Elsevier, License Number 5790741367977) [8].

The subsequent sections of this research work are structured as follows: Section 2 provides an in-depth survey of research endeavors aimed at identifying or predicting osteoporosis. Section 3 outlines the methodology proposed for this study. Section 4 discusses the experimental findings derived from the proposed approach, while Section 5 presents conclusions drawn from this work along with future research directions.

## 2. Related Work

In this segment, a review of literature was discussed concerning the diagnosis of osteoporosis utilizing artificial intelligence alongside other relevant methodologies. We delve into prevalent models leveraging X-ray imagery for the analysis of osteoporosis. SI is a straightforward technique for assessing bone mass using radiographs, has been the topic of several investigations, according to Hauschild et al. [9]. For all these trials, plain film radiographs were employed to assess SI. The major objective of their research was to assess SI gradings in digital radiography. Logistic regression analysis was performed to uncover salient determinants in the SI assessment process. Salamat, Mohammad et al., [10] compared the BMD measured with DXA to SI calculated using kappa statistics. They randomly selected 72 patients with neck or trochanteric fractures and enquired regarding their educational level, drug usage, gynaecological history, physical activity, and other lifestyle habits. They extracted the anterior-posterior radiographs to calculate were the SI. They assigned six SI grades using radiograph chat of the SI and compared to DXA results and discovered a substantial inter-observer variation and no link between SI and the skeletal mineral density assessment. They concluded that the index is unreliable and cannot be used for the clinical determination of osteoporosis.

Shankar et al. [11] measured the BMD of the right femur of 50 Indian women where none of the women had osteoporotic fractures. According to their study report 20% of Indian women have signs of being osteoporotic and 58% of postmenopausal women will have osteoporosis. They concluded that the SI was unreliable and had low utility in diagnosing femoral neck osteoporosis. The study lacked test cases as a limited magnitude of patients were in the trial, and the disease's risk factors were not considered.

In addition to computing the T and Z scores, Chen and Prasath [12] provided a unique approach that calculated individual scalar magnitude to indicate the interconnectivity of skeletal mineral constituents from DXA images. Their approach can be used to determine how well the bone components are linked. Rahul Paul et al. [13] used CNNs that were trained on ImageNet data to extract features and applied three different approaches to perform classification. Symmetric uncertainty selector was used to select 15 deep features and classified using random forest had better accuracy when compared to the other approaches used. Liu et al. [14] examined whether combining the Osteoporosis Self-Assessment Tool for Asians with SI is a better way to assess osteoporosis risk assessment within a sizable diabetic cohort. They concluded by stating that this tool together with SI could be a tool to screen hip fracture risk in diabetic population. Cruz et al. [15] assessed the primary methods that utilize artificial intelligence to identify populations susceptible to fractures or osteoporosis vulnerabilities. Their assessment indicates, several methods exist but they are specific to a particular age group or gender and ethnicity. They conclude by indicating that a research method must be developed to unify different databases and develop evidence-based standards for the evaluation of most significant risk factors.

Areeckal et al. [16] have suggested a low-cost automated technique to detect reduction in bone mass with the help of hand and wrist radiographs. Their pilot study show that the segmentation method used by them has high accuracy of detecting the region of interest of third metacarpal bone shaft and distal radius. They concluded by stating cortical radiogram metric measurements and cancellous texture analysis on the hand and wrist radiographs could be used for early detection of onset of osteoporosis. Lindsey et al. [17] engineered a deep neural network for the purpose of detecting and localizing fractures in radiographs. They used 135,409 radiographs to train it to accurately mimic the proficiency of 18 experienced sub-specialized orthopaedic surgical practitioners. They also conducted a controlled trial using emergency medical professionals to see how well could discern wrist radiograph fractures both with and without the application of the deep learning model. They concluded that medical practitioners in emergency when assisted by the automated system has reduced the diagnostic errors and enhanced the efficiency of the clinician. Fathima et al. [18] have proposed an algorithm to measure BMD and T-score using X-ray images. They have used X-ray bone images of different parts of the body like hand, spine, knee, ankle and clavicle. They pre-processed the images using Shock filter and segmented the pre-processed images using mean shift algorithm. They claim that the algorithm proposed by them produced better results.

Sato et al. [19] have created a computer-aided diagnostics system that achieves high accuracy; they have used the model EfficientNetB4 to construct a computer-aided diagnostic system for frontal plane hip radiographs, which were pre-trained on ImageNet. The dataset was created using 10484 medical images captured from three hospitals, annotated, labelled, and pre-processed. Their computer-aided diagnostics system extracted features from the points where the fracture was located and was visualized using Grad-CAM. They claim to have been able to attain an accuracy of around 96% and specificity, F-value and area under the curve of 96.9%, 0.961 and 0.99 respectively.

Erjiang et al. [20] carried out a study on a cohort of 13,577 Irish men and women aged above 40 and established that osteoporosis Self-assessment Tool Index played an important role to classify for osteoporosis using BMD. They also performed comparative analysis with seven distinct ML techniques could improve DXA detection of osteoporosis. Shim et al. [21] worked on a dataset that contained nineteen features of 1792 postmenopausal Korean women, 613 among them had osteoporosis. They applied a feature selection method to reduce the original 19 features to nine features and investigated seven machine learning models to forecast the risk of osteoporosis. Their outcomes demonstrate that conducting the experiment with both original and extracted features yielded comparable or superior results. They also claim that artificial neural network (ANN) outperformed other models as it exhibited the highest area under the receiver operating characteristic curve value. Dildar Hussian et al. [22] reported a technique based on deep learning for femur segmentation in DXA imaging, intending to enhance segmentation accuracy. The results show that the Fully Convolutional Network model may be used to segment DXA images since it results in higher accuracy and sensitivity on femur DXA images when properly tuned. They tried transfer learning to increase the capability of the Fully Convolutional Network model. The dataset used was 600 femur images segmented by radiology experts. Yamamoto et al. [23] conducted a study to evaluate whether integrating clinical data improves the diagnostic performance of deep learning in comparison to solely relying on images. They investigated the performance of five models and concluded that the EfficientNetb3 the highest performance. Zang et al. [24] conducted a multicenter retrospective cohort study to construct a deep CNN model. Based on the DXA, BMD, and TS, they divided the patients into three groups and used details of Chinese females aged 20-40 as a reference for calculating T-scores.

Tecle et al. [25] used 4000 posteroanterior radiograph images collected from the health Lab of Rochester University to study the effect of using second metacarpal cortical percentage (2MCP) to predict osteoporosis. They developed a network that segmented the second metacarpal prior to which they standardised the images by performing laterality correction and vertical alignment correction. Manual measurement was done on the images by a trained computer expert and validated by a hand surgeon, the middle third of the 2MCP was used for the purpose of prediction. The model used for prediction was based on AlexNet the pre-trained model was further refined using their training data and tested which yielded an accuracy of 88.4%. Anam et al. [26] conducted a comprehensive survey on the utilization of trabecular bone in osteoporosis research. They did an exclusive search which resulted in listing 1050 articles of which on eliminating duplication they came up with 533 articles for reviewing. They investigated the abstract and subsequently screened them which led to 62 studies as the final list. On reviewing the 62 papers they concluded that majority of the articles use support vector machine and 3D-micro-MR imaging. Ryoungwoo Jang et al. [27] applied DL algorithms to forecast osteoporosis based on hip radiographs. They developed a deep neural network (DNN) model derived from the VGG16 architecture, enhanced with a nonlocal neural network component, for this purpose. The visualization of model performance was achieved through the overlay of a gradient-based classifier map onto the original image. They claim that the DNN model proposed by them is highly effective screening tool for osteoporosis prediction as they were able to attain improved accuracy, sensitivity, and specificity.

Karen Simonyan and Andrew Zisserman [28] explored the influence of network depth on image classification accuracy. According to their findings, increasing the depth of the conventional convolutional network architecture positively affects classification accuracy. Huang Gao et al. [29] introduced a Dense Convolutional Network (DenseNet) that connects each layer to all the subsequent layers in a feed forward manner. Their experiment results show that the network shows consistent improvement in accuracy even with growing number of parameters and does not affect performance. They claim that DenseNet achieves performance at par with existing methods that too with lesser computation time and few parameters. Kaiming et al. [30] proposed a residual learning framework that ease the training the networks which are very deep. They claim that the residual network proposed by them are easier to optimize and gain accuracy when the depth is increased considerably. Jimmy Ba et al. [31] suggested that normalizing neuron activities could decrease training time. They introduced a normalization method called "layer normalization," which is a variation of batch normalization. They assert that their normalization approach accelerates training compared to the baseline model, although batch normalization achieves better performance.

Kingma and Ba [32] introduced an effective algorithm for computing gradient-based stochastic objective functions. Their algorithm purportedly combines the strengths of two widely used methods: AdaGrad and RMSProp. They argue that their approach is robust and suitable for addressing non-convex optimization challenges in machine learning. optimization in optimizing some scalar parameterised objective function requiring maximisation or minimization concerning its parameters [32]. Krizhevsky et al. [33] have proposed a large deep CNN. The network proposed by them handled 60 million parameters and had 6.5 lakhs neurons spread across five convolutional layers, 3 max pooling layers and three fully connected layers. Finally, 1000 class classification was done using SoftMax. They have used non-saturating neurons and efficient graphical processing unit's implementation to ensure faster training and were able to achieve top-5 error rate of 17%. They also tried a variant and able to reduce the top-5 error rate to 15.3%.

Vinod and Geoferry [34] used a type of hidden unit for a restricted Boltzmann machines that tied up the weights and biases of an infinite set of binary units. They approximated the stepped sigmoid units with noise rectified linear units and witnessed better results for problems of object recognition and comparing faces. Sebro and Garza-Ramos [35] in their study introduced an artificial intelligence (AI)-based methodology designed for the automated detection of osteoporosis through the analysis of X-ray images. Their research endeavor involved the formulation of a diagnostic model for osteoporosis utilizing handcrafted features derived from descriptors acquired through an extensive analysis of bone images. Notably, two established handcrafted methods, namely Histogram of Oriented Gradients and Local Phase Quantization, were employed for the extraction of descriptor features. In pursuit of optimal performance, a bat-based optimization method was adopted to discern suitable parameters for the Gabor filters. Furthermore, information integration transpired at two distinct levels, encompassing fusion at the score level and fusion at the decision level. It also found that the proposed AI-based technique achieved high accuracy in detecting osteoporosis through X-ray image analysis. Handcrafted features and Gabor filters with different orientations significantly improved the model's accuracy. The bat-based optimization method also played a crucial role in determining appropriate filter parameters.

Meriem Mebarkia et al. [36] compared digital breast tomosynthesis and digital mammography for breast cancer detection in 1,000 patients without prior breast cancer. Digital breast tomosynthesis outperformed digital mammography by exhibiting higher diagnostic accuracy (92.5% vs. 87.5%), sensitivity (92.3% vs. 84.6%), specificity (92.6% vs. 88.1%), and positive predictive value (70.6% vs. 56.3%). A two-step DL approach was employed YOLOv4 samples and Unet models for the detection and segmentation of linear microcracks within trabecular bone [37]. The study aims to improve the accuracy and efficiency of microdamage evaluation in bone samples, which is currently a time-consuming and subjective process.

The potential of computer vision and ML techniques were explored to enhance the identification of individuals at risk of fractures beyond the capabilities provided by clinical risk factors and BMD measurements obtained through DXA, using HR-pQCT scanning [38]. The investigation employed a random forest classifier to categorize participants based on their history of previous fractures. The study concludes that ML techniques, combined with HR-pQCT imaging and clinical variables, could improve the identification of those at risk of fracture. Lu et al. [39] reported their work whose primary objective was to cultivate DL algorithms utilizing lateral spine X-rays to streamline the identification of osteoporosis. The prevalence of vertebral fracture was 17.1%. In a subset of individuals with DXA measurement, 38.2% had osteoporosis. In the test set, the area under the receiver operating characteristics of DNN models to detect prevalent vertebral fracture and osteoporosis was 0.94 and 0.86, respectively.

Hong et al. [40] developed a deep learning (DL) algorithm to estimate the native hip joint center from standard pelvis radiographs. They assert its ability to reduce variability and subjectivity in hip joint center estimation, using the femoral head center as a reference. However, they highlight the need for extensive external validation through clinical cohort studies to ensure its operational safety. The study utilized data from the Osteoarthritis Initiative, housed in the National Institute of Mental Health data archive, for instrument preparation. Their method presents potential for improving hip joint center estimation, carrying significant implications for clinical practice and hip joint research.

Jang et al. [41] provided a useful manual for creating, implementing, and designing DL pipelines for musculoskeletal imaging that could be used by orthopaedic physicians. The guide covers all the important phases in DL projects, such as problem definition, team building, data acquisition and curation, labelling, ground truth establishment, pre-processing and data augmentation, hardware selection, training and assessment of DL models, model architecture selection, hyperparameter optimization for fine-tuning, utilizing pre-trained models and transfer learning, choosing a loss function, and assessing model performance.

Jacob et al. [42] propose a DL classification model designed to utilize 2D projection images from CT data or plain radiographs for predicting the likelihood of hip fracture. The method they introduce is asserted to be entirely automated, making it well-suited for opportunistic screening environments. This capability could facilitate the identification of high-risk individuals within a larger population without necessitating additional screening procedures. According to the authors' conclusions, the DL model might find application in opportunistic screening setups to detect high-risk individuals without additional screening. Additionally, it holds promise for improving hip fracture risk prediction in males and could serve as a complementary tool to DXA scans in diagnosing osteoporosis.

Schmarje et al. [43] explored challenges in existing linear inversion algorithms used in ultrasound CT bone imaging. They discussed impediments like significant impedance contrast and periosteal interface effects, hindering high-resolution image delivery. Despite lengthy computation times making them impractical clinically, the study proposes non-linear approaches as potential solutions. They introduced a rapid-acquisition setup using classical Born approximation and spatial Fourier transform for quicker acquisition and low-intensity sonication. Though current results showed suboptimal contrast-to-noise ratio, the paper anticipates non-linear methods to improve future clinical applications in bone imaging via ultrasound CT.

Fradi Marwa et al. [44] conducted a study investigating the utilization of ML algorithms in predicting pharmacotherapy outcomes for osteoporosis. They compared the performance of four different algorithms and evaluated them using receiver operating characteristic curves. The aim of the study was to establish a clinical approach for personalized care. The findings of the study indicate that ML algorithms may facilitate the development of a therapeutic strategy tailored to individualized treatment of osteoporosis. These algorithms can identify crucial input features associated with treatment outcomes and predicting the likelihood of treatment success for individual patients. Consequently, the study suggests that ML algorithms hold the potential to enhance the efficacy of pharmacotherapy for osteoporosis. Yi-Ting Lin et al. [45] investigated the utilization of DL for extracting fracture risk from DXA images and detecting pathological characteristics. A CNN was trained by the authors on a dataset of DXA images, and its performance was assessed across three classification tasks: lumbar scoliosis identification, fracture risk prediction, and differentiation between normal and abnormal DXA images.

Nissinen et al. [46] conducted a comprehensive review of recent advancements in automatic classification and grading techniques for assessing knee osteoarthritis severity using X-ray images. The authors compared different machine learning-based methodologies, encompassing transfer learning, active learning, and DL, against conventional manual techniques. Additionally, they deliberated on the obstacles and prospective avenues for automatic knee osteoarthritis severity classification. Nevertheless, the authors concluded that additional research was required to tackle the challenges and constraints of these methodologies and to formulate more resilient and broadly applicable models.

Deepak et al. [47] developed a robust combined DL and ML approach for automatically detecting hip prosthetic loosening. The approach involved extracting features from X-ray images, which were subsequently analyzed using support vector machine, random forest, and principal component analysis. Through SHapley Additive exPlanations analysis, two features not inherent to the original image were identified, while the most influential characteristics were determined to be part of the original image. The proposed method was suggested as a viable tool for identifying implant failure in patients undergoing hip replacement follow-up.

Muscato et al. [48] found that incorporating clinical and 2D patient-specific biomechanical data into an ML classifier through a multi-technique approach could be utilized to model and improve hip fracture prediction. Following a comprehensive validation process that assessed numerous classification techniques, support vector machine was identified as the most proficient model. The model yielded a 14% increase in accuracy compared to the gold-standard BMD. Additionally, it was observed that integrating biomechanical and clinical data was a swift, straightforward, and cost-effective method for incorporation into daily clinical practices.

Villamor et al. [49] presented a DL method capable of analyzing standard Full-limb radiographs for knee phenotypic classification and alignment. They observed that knee alignment was influenced by sex and Kellgren-Lawrence (KL) grade, noting that with higher KL grades, more knees exhibited varus alignment. The most prevalent knee phenotype for KL grades 0 through 3 was type II knees (neutral, apex distal Joint Line Orientation), whereas for KL grade 4 knees, the most prevalent knee phenotype was type I knees (varus, apex distal Joint

Line Orientation). This method may be reliably utilized in populations of patients with arthritis to determine constitutional lower limb alignment.

The existing literature shows the predominant use of DXA images for the measurement of BMD to assess the osteoporosis status of the patients. This is done by using the measures of T and Z scores. Various automatic and semi-automatic artificial intelligence and machine learning-based techniques are proposed and applied successfully to aid the analysis of the DXA images. On the other hand, SI based diagnosis proves to be a cost-effective technique used manually to categorize the X-ray images into 6 different grades for osteoporosis classification. Only few studies compare the usage of both the DXA and SI techniques for osteoporosis diagnosis.

From the comprehensive literature, the application of ML techniques for classifying these 6 grades of SI from unlabelled hip X-ray images is not available. Developing such techniques can be useful in analyzing osteoporosis in an early stage in a much cheaper way. It can also help physicians as a screening tool that can be used in early diagnosis. Hence, this study uses CNNs to extract the features from the unlabelled sample of 1597 pelvic X-ray images and cluster them based on SI grades. Based on the cluster type, an assessment of osteoporosis can be done.

The major contributions of this study are as follows.

- While existing research predominantly focuses on using DXA images and measures like T and Z scores for assessing osteoporosis, there is a lack of studies comparing DXA and SI techniques. This study fills this research gap by exploring the application of machine learning (ML) techniques, specifically convolutional neural network (CNN), for classifying six grades of SI from unlabelled hip X-ray images.
- The study introduces ML techniques, particularly CNNs, for categorizing X-ray images based on SI grades. This approach offers a cost-effective alternative to traditional DXA imaging, potentially facilitating early-stage osteoporosis detection in a more affordable manner.
- By employing ML methods to cluster pelvic X-ray images according to SI grades, the study aims to provide a screening tool for early osteoporosis diagnosis. This has implications for improving patient care by enabling timely intervention and management of osteoporosis, ultimately enhancing overall health outcomes.

### 3. Proposed Model for Feature Extraction

CNNs like DenseNet, InceptionV3, and EfficientNet are among the first used to extract features from the X-ray images. When K-Means clustering is used, these networks, however, produce low Silhouette Scores, which point to less-than-ideal outcomes. As a result, a unique CNN architecture is developed, which consists of four convolutional layers with 64, 64, 128, and 128 neurons each, then 2D max pooling. The activation function that is used for each convolutional layer is a rectified linear unit, which is followed by a dropout of 0.5. The flattened features are then run through two dense layers, each with 64 and 16 nodes. Several clustering algorithms are then applied to the extracted features to determine the ideal number of clusters and extract insights from the data after feature extraction.

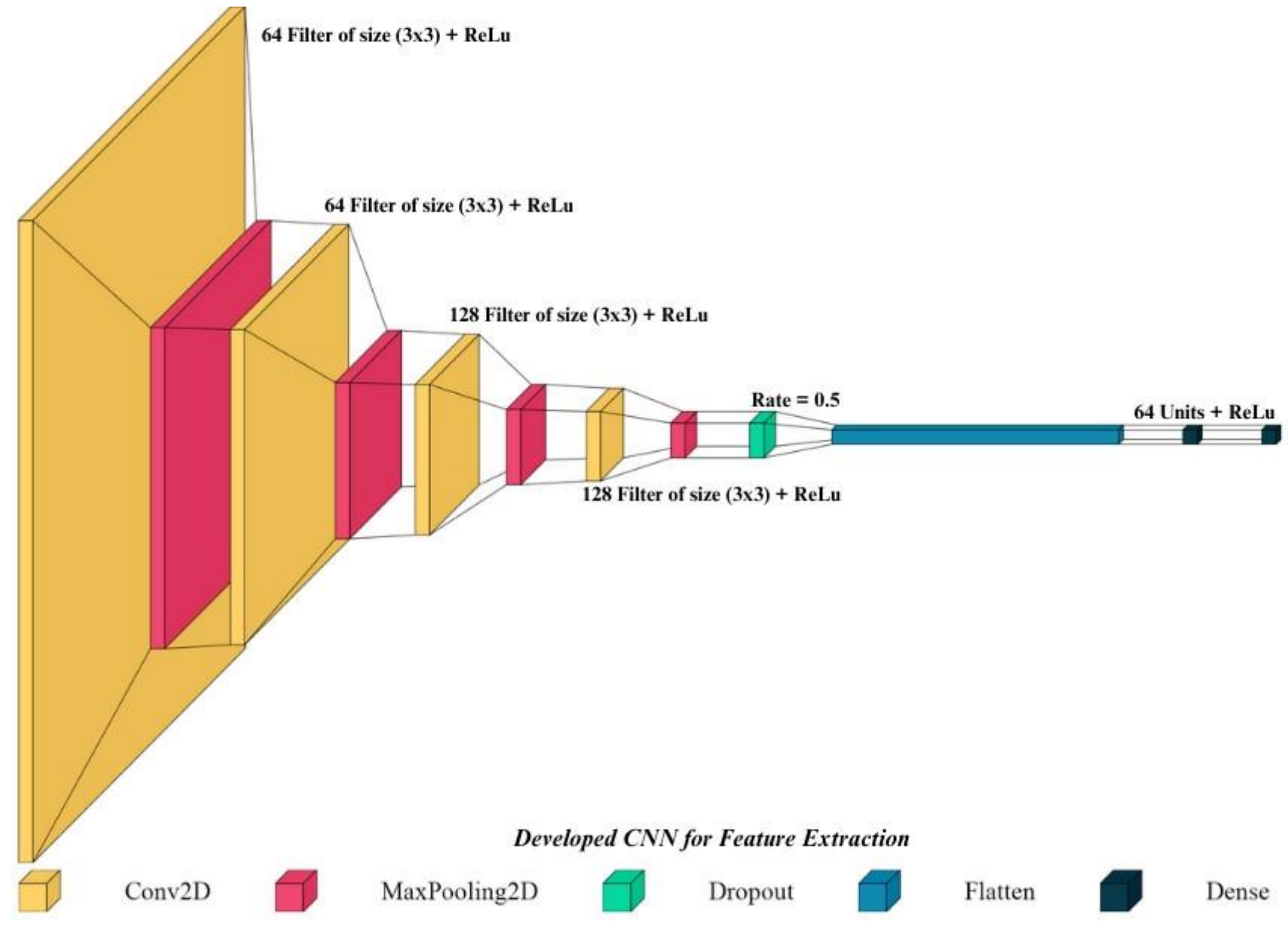

**Figure 2.** Architecture of Custom CNN.

### 3. Methodology

The execution process of this study is depicted in Figure 3. The overall execution is split into different processes such as data collection, data pre-processing, Feature extraction and clustering and finally the measurement and analysis of the Silhouette score. There processes are elaboration in the sections that follow.

*3.1 Data Collection*

The X-ray images are collected from Medical College Hospital and Research Centre. The images from the Department of Radiology recorded from Jan 2021 to March 2022 were taken for this work. To maintain the patients' privacy, age and sex were the only parameters taken along with the images. The dataset acquired consists of 838 hip X-ray images collected from male and female patients aged between 20 and 70. The images used in this work are unlabelled. Hence, Unsupervised ML approach was adopted to identify the various clusters to which the X-rays belong that would map to the SI. Figure 4 shows the six grades of the SI from the X-ray images.

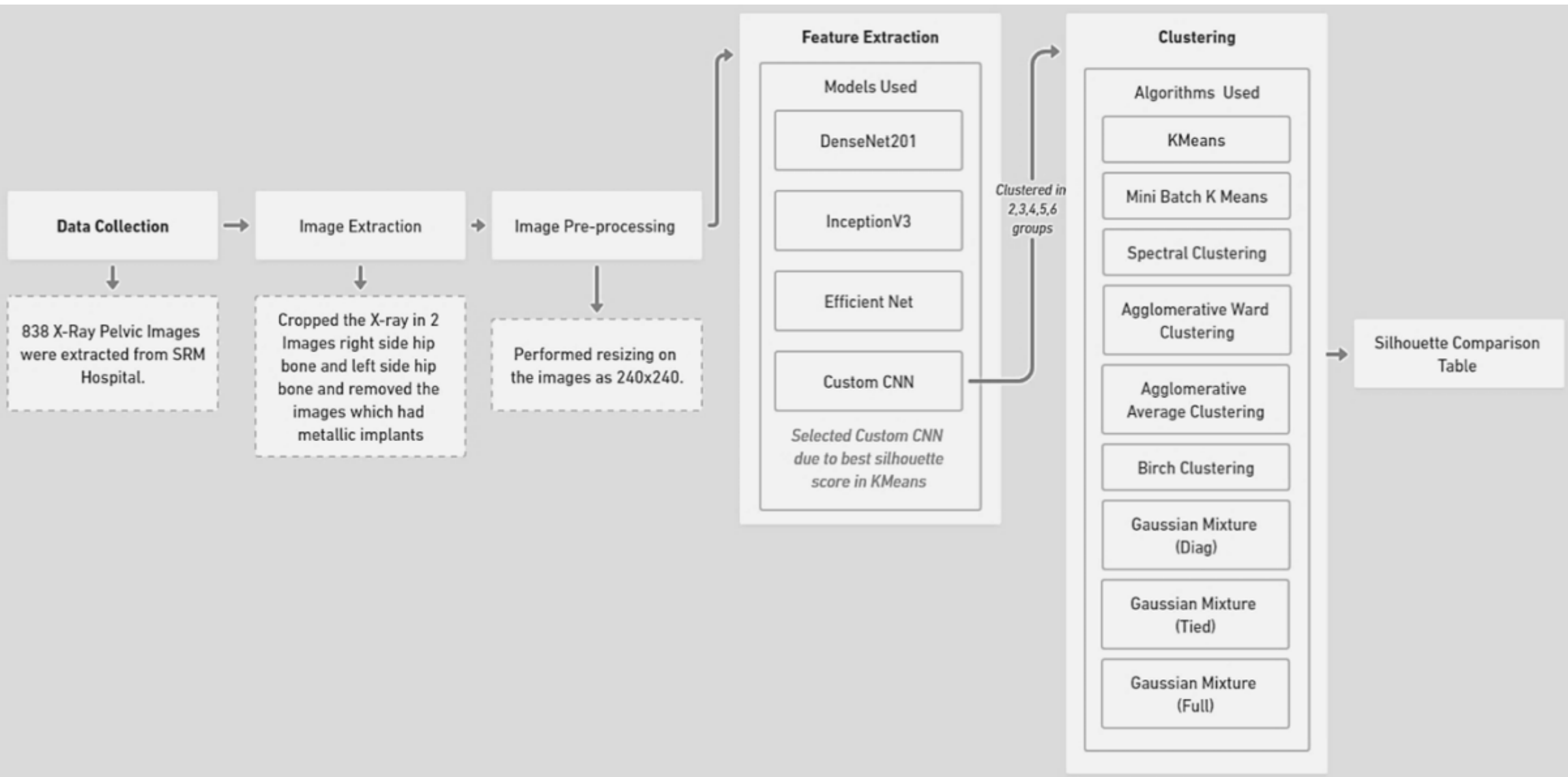

**Figure 3.** Overall process flow

*3.2 Data Pre-processing*

The images were then cropped to ensure that the region of pelvic bone where the osteoporotic patterns are visible. All the images were then resized to the same size to get an accurate result while performing feature extraction using advanced CNNs like DenseNet, InceptionV3, and EfficientNetB0. Before performing feature extraction, the images were converted into a NumPy array encoding for a batch of images. After utilizing transfer learning to extract features, each image feature was fitted into K-Means clustering to construct clusters based on the features recovered by the CNNs.

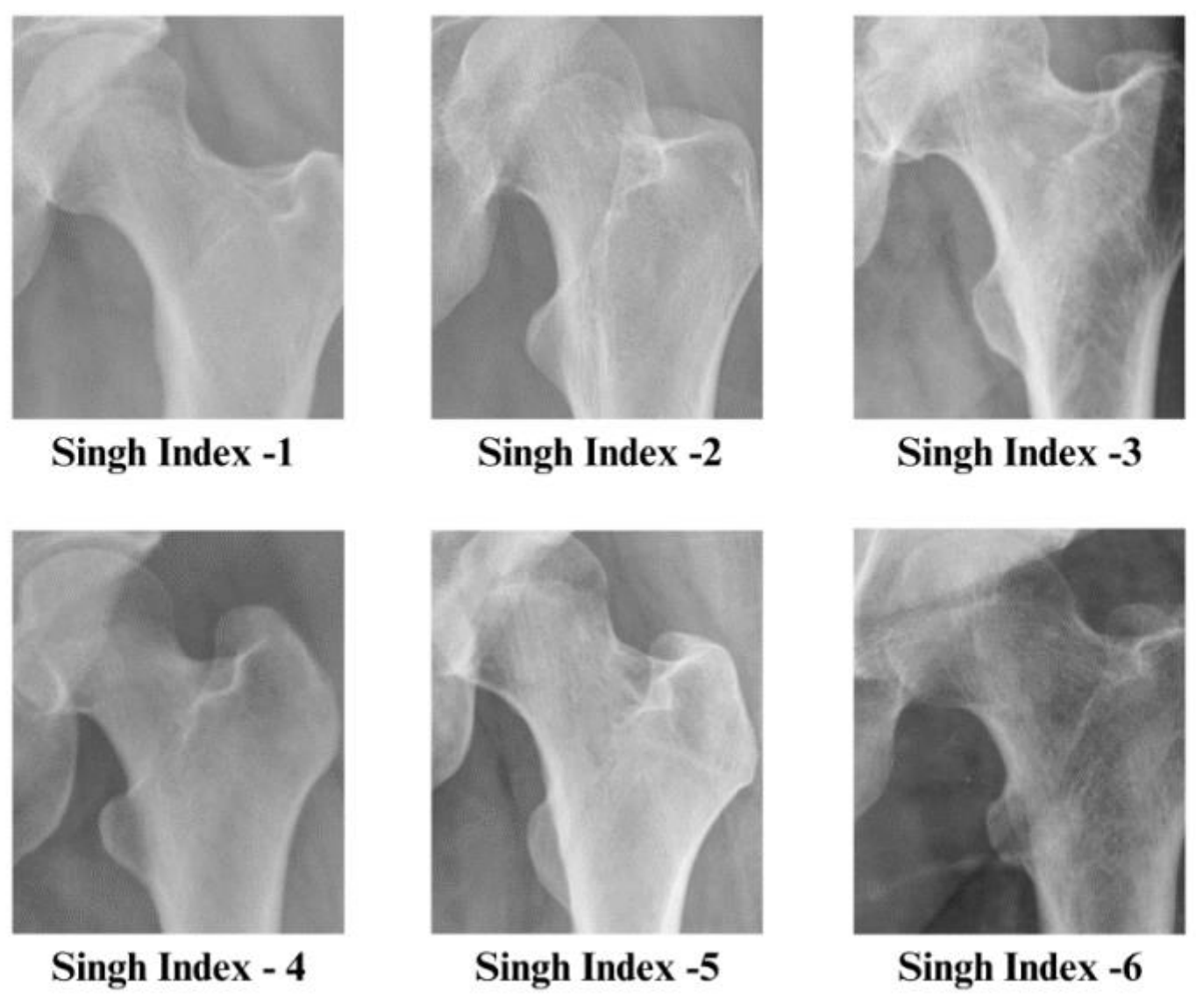

**Figure 4.** Reference X-rays with SI

*3.3 Feature Extraction*

The feature extraction is done using the transfer learning technique using CNNs pre-trained with an ImageNet dataset.The advantages of ImageNet for feature extraction for pre-trained CNNs include its enormous scale, diverse class distribution, emphasis on visual similarity, transferability, generalizability, and localization capabilities. These characteristics promote the creation of powerful, precise, and highly adaptive features that perform well in a variety of subsequent computer vision tasks.Three different CNN architectures are used in this study, namely DenseNet-201, InceptionV3, and EfficientNetB0. The architecture of InceptionV3 has several improvements, such as label smoothing, factorise 7 x 7 convolutions, use of an auxiliary classifier, etc. [50]. A DenseNet-201 is a 201 layers-deep network pre-trained with more than a million images and can classify the images into 1000 classes. Each layer collects additional inputs from all preceding levels and passes on its feature maps to all following layers to maintain the feed forward nature [51]. Unlike the current technique, which arbitrarily scales these components, EfficientNet uses a scaling method that uniformly scales all depth/width/resolution dimensions using compound coefficients. In addition to squeeze-and-excitation blocks, the base EfficientNetB0 network is built on the inverted bottleneck residual blocks of MobileNetV2 [52, 53]. To perform transfer learning and extract the features from the images, the TensorFlow library was used, which helps to predict the features of images. The images are converted into a tensor using TensorFlow, and pre-processing is done. Once the features are extracted from these images, they are passed onto the next step, clustering, to be classified.

*3.4 Clustering*

The clustering of X-ray images into six groups corresponding to SI grades has been accomplished through the application of the K-Means clustering algorithm. K-Means, a prominent technique in Unsupervised ML, serves as a straightforward and efficient clustering algorithm. In the context of this study, the algorithm is employed to establish six groups aligned with specific criteria related to SI. Emphasizing the significance of cluster performance and error criterion within the K-Means algorithm, the objective is to minimize these criteria. The algorithm initiates by selecting focal points to represent the clusters, and the first K sample dots serve as the initial cluster focal points. Subsequently, the remaining dots are assigned to their respective focal points based on the minimum distance criterion, resulting in the initial classification aligned with different grades of SI. Figure 5 showcase the intricate processes of feature extraction and clustering, employing diverse CNNs. The work by Ezugwu et al. [54] serves as a valuable reference for the state-of-the-art in clustering algorithms and ML applications.

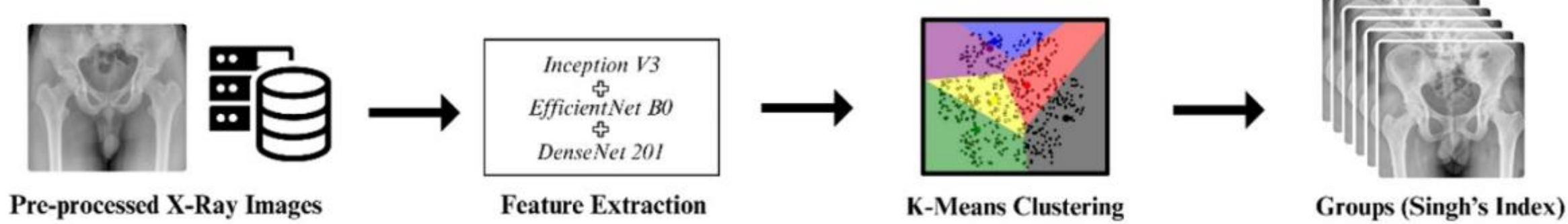

**Figure 5.** Feature extraction and clustering

Drawing insights from the work of Szegedy et al. [50], the InceptionV3 model designed for image recognition is structured with a combination of symmetric and asymmetric building blocks. A notable characteristic of this architecture is the extensive application of batch normalization, and the computation of loss is carried out through the SoftMax function. Leveraging the insights from Huang et al. [28], DenseNet adopts a unique approach by concatenating all preceding outputs as input for subsequent layers. This methodology addresses the vanishing gradient problem inherent in architectures with long distances between input and output layers, contributing to enhanced accuracy in challenging scenarios. In a distinct vein, as proposed by Tan and Le [51], EfficientNet introduces a principled scaling methodology characterized by uniform scaling across all dimensions. This scaling is achieved through a compound coefficient, demonstrating a thoughtful and efficient strategy for model scaling in CNNs.

In this study, while InceptionV3, DenseNet, and EfficientNet were initially considered for feature extraction, a simple CNN was employed due to its superior performance and efficacy in achieving the desired results. This decision was based on the outcomes of our experimentation, emphasizing the importance of tailoring the feature extraction and was reinforced by the results obtained in clustering analysis, where it demonstrated superior performance and yielded improved Silhouette Scores.

### *3.5 Comparison of Algorithms*

Clustering groups similar data points based on their features. This research aims to group the SI of Osteoporosis for the femoral bone X-ray image using the extracted features. Various clustering algorithms have been used, such as K-Means, DBSCAN, and agglomerative clustering. This work uses unsupervised methods for the classification purpose. Hence, comparing different clustering algorithms can provide a much greater insight into the problem and the dataset used. The following subsections give the description about the clustering algorithms used for comparison in this experiment.

#### *3.5.1 K-Means Algorithm*

The feature extraction outcomes are subjected to the K-Means algorithm, a widely adopted clustering technique [56]. Assigning each data point to the closest centroid, the method starts by randomly initializing K centroids. Then, by calculating the mean of every data point assigned to them, the centroids are updated. The centroids stabilize at some point during the iterative process, or a predefined number of iterations is reached. The objective is to minimize the sum of squared distances between each data point and its designated centroid [54].

### *3.5.2 Mini Batch K-Means Algorithm*

The Mini Batch K-Means technique is a variation of the K-Means algorithm that uses a random selection of the data instead of the entire dataset to update the centroids in each iteration. By doing this, the process is accelerated without sacrificing the quality of the clusters produced. In this work, the Mini Batch K-Means algorithm was applied to the extracted features of the CNN model.

### *3.5.3 Spectral Clustering Algorithm*

The clustering method known as spectral clustering divides the data into groups based on the eigenvalues and eigenvectors of the similarity matrix [57]. The similarity matrix is built based on how similar the data points are. The algorithm located the similarity matrix's K largest eigenvectors, which then applies K-Means clustering to the resulting matrix. In this study, the retrieved features from the CNN model were subjected to the spectral clustering algorithm. The technique searches for the K largest eigenvectors using the similarity matrix created from the extracted features. Then, it clusters the generated matrix using K-Means.

### *3.5.4 Agglomerative Clustering Algorithm*

The agglomerative clustering technique is employed, creating a single cluster by iteratively merging related data points or clusters [58]. Once a preset number of clusters is reached, the algorithm iteratively merges the two closest clusters starting with each data point as a single cluster. This study employed two iterations of the agglomerative clustering technique: average and Ward. The average variation seeks to reduce the average distance between data points in each cluster, whereas the Ward variant seeks to minimize the sum of squared distances inside each cluster. The retrieved features from the CNN model were subjected to both approaches.

### *3.5.5 Birch Clustering Algorithm*

A clustering algorithm called Birch Clustering divides the data into clusters using a hierarchical method [59]. The clustering feature tree, a tree-like data structure that the algorithm first creates by clustering the data points according to a predetermined threshold, is the result of this process. Extracted features from the CNN model were then employed in the Birch Clustering algorithm, which constructs the clustering feature tree with the extracted features and performs clustering using K-Means.

### *3.5.6 Gaussian Mixture Model*

A probabilistic clustering technique known as the Gaussian Mixture model posits that the data originates from a synthesis of Gaussian distributions [60]. It calculates the mean and covariance of each distribution and determines which k-Gaussian distributions suit the data the best. The probability that each data point will belong to a certain Gaussian distribution is then determined based on the likelihood that the data point was produced from that distribution. The algorithm then gives each data point the highest probability Gaussian distribution. It is superior to conventional clustering algorithms in several ways, including its capacity to handle clusters of various forms and sizes and its use of soft clustering [54]. However, it requires a lot of calculation and might reach a local optimum.

## 4. Results and Discussion

The analysis and interpretation of the results are discussed in detail in this section. The distribution of images across different clusters can provide some insights into the underlying dataset's characteristics. To make a conclusive judgement about the performance of the networks in terms of clustering, we need to consider the clustering evaluation metric. For this purpose, the Silhouette score metrics is proposed to be used, which will give the goodness of the number of clusters for the sample dataset used. A higher value of this score indicates good fit of the clusters. Table 1 gives these Silhouette score values for different clustering algorithms used in this study. It can be inferred that the K-Means algorithm performs well with 2 clusters, with a Silhouette score of 0.8097. However, as the number of clusters increases, the Silhouette score decreases gradually. This indicates that the K-Means algorithm may not be the best choice for clustering with more than 2 clusters. Similarly, the Mini Batch K-Means algorithm performs well with 2 clusters, with a Silhouette score of 0.8691. However, as the number of clusters increases, the Silhouette score decreases rapidly, indicating that the Mini Batch K-Means algorithm may not be suitable for clustering with more than 2 clusters. The spectral clustering algorithm performs well with 2 and 3 clusters, with Silhouette Scores of 0.8515 and 0.7821, respectively. As the number of clusters increases, the Silhouette score decreases gradually. This indicates that the spectral clustering algorithm may be suitable for clustering with up to 3 clusters.

**Table 1.** Silhouette score of X-rays grouped into different grades by various algorithms.

| Algorithm | 2 Clusters | 3 Clusters | 4 Clusters | 5 clusters | 6 clusters |
|---|---|---|---|---|---|
| K-Means | 0.8097 | 0.7428 | 0.7486 | 0.70346 | 0.6560 |
| Mini batch K-means | 0.8691 | 0.6370 | 0.4970 | 0.2233 | 0.3877 |
| Spectral clustering | 0.8515 | 0.7821 | 0.7621 | 0.7648 | 0.7102 |
| Agglomerative Ward clustering | 0.7826 | 0.7760 | 0.7779 | 0.7372 | 0.6232 |
| Agglomerative average clustering | 0.9360 | 0.9164 | 0.8681 | 0.8640 | 0.8565 |
| Birch clustering | 0.9234 | 0.9164 | 0.8683 | 0.8709 | 0.8662 |
| Gaussian mixture (Tied) | 0.8273 | 0.7903 | 0.7611 | 0.7612 | 0.7164 |
| Gaussian mixture (Diag) | 0.6935 | 0.6081 | 0.4727 | 0.4286 | 0.4144 |
| Gaussian mixture (Full) | 0.7022 | 0.5967 | 0.5856 | 0.5106 | 0.5474 |

The agglomerative clustering algorithm has two variations, Ward and average. The Ward variant performs slightly better than the average variant in this case. The agglomerative clustering algorithm performs well with 2 and 3 clusters, with Silhouette Scores of 0.7826 and 0.7760, respectively. As the number of clusters increases, the Silhouette score again decreases gradually. This indicates that the agglomerative clustering algorithm may be suitable for clustering with up to 3 clusters. In the case of the Birch Clustering algorithm, it performs well with 2 clusters, with a Silhouette score of 0.9234. However, as the number of clusters increases, the Silhouette score decreases gradually. This indicates that the Birch Clustering algorithm may not be the best choice for clustering with more than 2 clusters. The algorithm has three variations: Tied, Diag, and Full. The Tied and Diag variants perform similarly, while the Full variant performs slightly worse. The Gaussian Mixture algorithm performs well with 2 clusters, with Silhouette Scores of 0.8273 and 0.6935 for the Tied and Diag variants, respectively. However, as the number of clusters increases, the Silhouette score decreases rapidly, indicating that the Gaussian Mixture algorithm may not be suitable for clustering with more than 2 clusters. Figure 6 shows a graphical representation of the Silhouette Scores obtained for different algorithms. From this chart, we can find that most of the X-ray images of the dataset are normal images without definite osteoporosis. A random check on the datasets found that the ratio of normal images is higher than the osteoporotic X-rays. The results achieved from the study confirm the same.

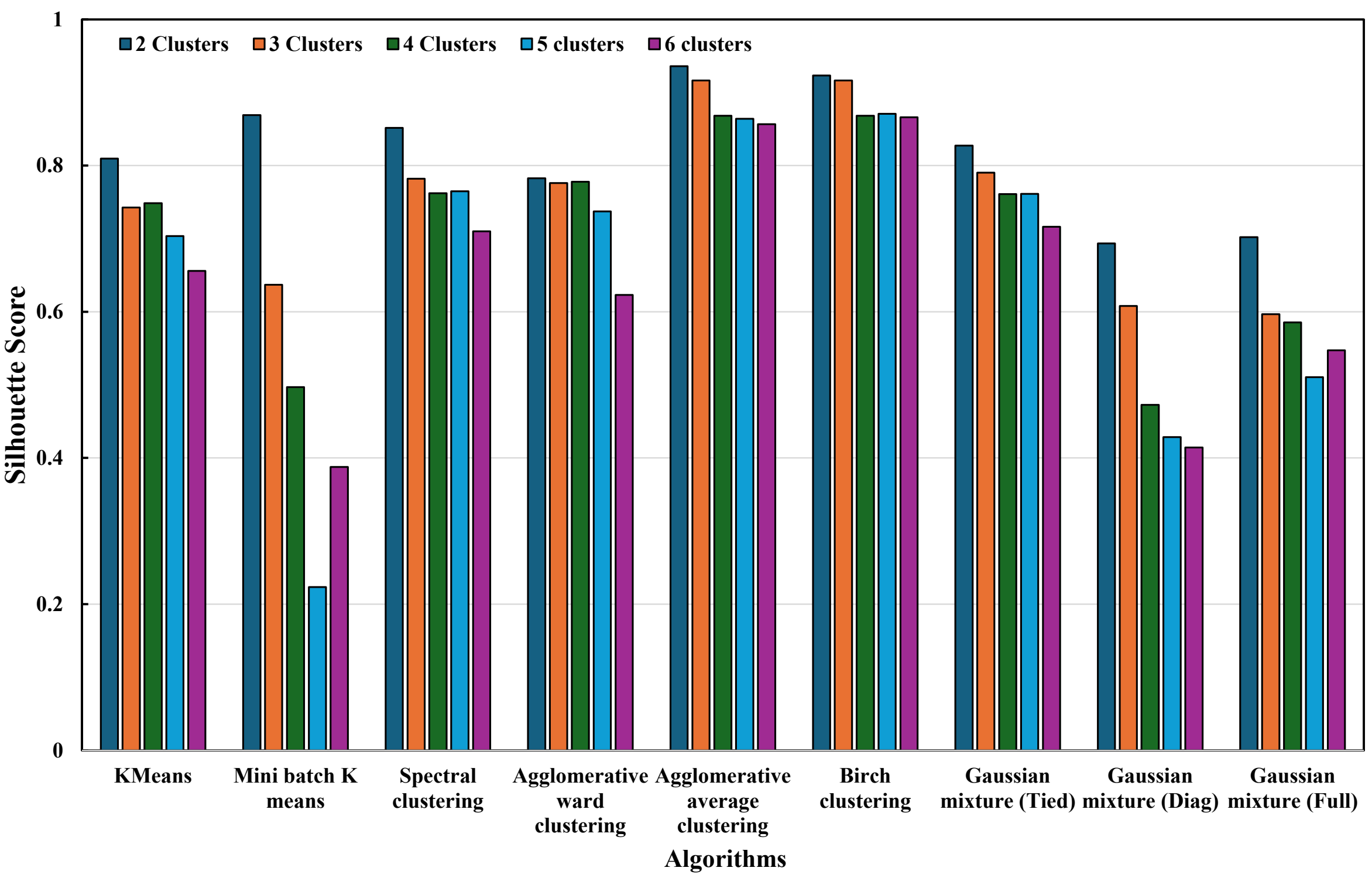

**Figure 6.** Silhouette score of X-rays vs Clusters by Various Algorithms.

There are also some variations concerning the classification done by different CNN architectures due to the following reasons:

- Data being unbalanced – Even Though the dataset consists of an ample 838 images, the ratio of the images in each category is not the same. The manual analysis of random samples of the data by the expert shows that the images belonging to grade 4 and above are more than those under grade 3 and below.

- Image Quality for clustering - The intensity of images is pivotal for effective feature learning in neural networks. In this study, image pre-processing involves cropping and resizing, crucial for subsequent clustering analyzes. Additional steps may include the application of filters to augment trabeculae visibility and overall image quality and optimizing the input for clustering algorithms.
- Information used for prediction - In this study, the sole utilization of X-ray images as input for osteoporosis prediction is acknowledged. However, practical considerations suggest the necessity of incorporating additional patient metadata, including sex, age, and pertinent clinical factors. This multi-modal approach would enhance the predictive model by incorporating diverse data sources, providing a more comprehensive foundation for accurate predictions.
- Sample reference data - Due to the unavailability of labelled data, this study used an Unsupervised ML technique to develop the prediction model. Instead, Semi-supervised and self-supervised learning techniques may provide improved results when combined with other dependent variables, as discussed earlier, can provide usable results.

## 5. Conclusion

The present study aimed to conduct an extensive performance analysis on the application of ML models for predicting SI using pelvic X-ray images. Rather than utilizing DXA images, which are routinely employed to diagnose osteoporosis by calculating BMD, this investigation opted for X-ray images. The objective was to develop an automated system employing ML algorithms that could prove beneficial in settings where DXA imaging machines might not be available. The study aimed to analyze the feasibility of utilizing unlabelled pelvic X-ray images to cluster them according to SI grades, without incorporating additional clinical parameters. The Silhouette score was employed to gauge the suitability of different cluster numbers for the dataset. Based on the Silhouette Scores, spectral clustering and agglomerative clustering algorithms performed well with up to 3 clusters. Conversely, K-Means, Mini Batch K-Means, Birch Clustering, and Gaussian Mixture algorithms performed effectively with 2 clusters but might not be optimal choices for clustering beyond 2 clusters. Initially, an endeavor was made to classify X-ray images into 6 clusters corresponding to the six SI indices. However, feedback from medical practitioners indicated that the preliminary model proposed did not achieve the expected cluster classification. This was attributed to several factors, including dataset imbalance, dependency on image quality, lack of additional clinical patient information, and the absence of labels for the data.

To address these challenges, it is proposed to utilize reference images representing centroids for each SI grade and cluster the images to assess model performance. However, there is a possibility that the images may not be evenly distributed to represent all SI grades. Additionally, it is suggested to annotate the images with the assistance of domain experts and explore the use of semi-supervised and self-supervised learning algorithms in the future. Moreover, incorporating clinical parameters alongside X-ray images could enhance model performance.

**Competing interests**

The authors declare no potential conflict of interest.

**CRediT Authors Statement**

**Vimaladevi Madhivanan:** Conceptualization, Formal analysis, Methodology, Investigation, Writing–review & editing, **Kalavakonda Vijaya:** Conceptualization, Formal analysis, Methodology, Investigation, Writing–review & editing. **Abhay Lal:** Curation, Methodology, Investigation, Data visualization, Data validation, Original Draft-writing, **Senthil Rithika:** Formal analysis, Methodology, Investigation, Original Draft-writing, **Shamala Karupusamy Subramaniam:** Data validation, Data visualization, Writing–review & editing, Mohamed Sameer: Investigation, Data visualization, Data validation, Writing–review & editing. All authors have read and agreed to the submitted version of the manuscript.

**Funding information**

No funding was received.

**Availability of data and materials**

The data will be available on request.

**Acknowledgements**

We would like to thank Radiology Department, SRM Medical College Hospital and Research Centre, for supporting us with the requested number of X-ray images. We appreciate the support provided by Salwyn Joseph Mathew and Ayushya Jain to retrieve the X-ray images.